\documentclass[aps,prl,twocolumn,superscriptaddress,nobibnotes,amsmath,amssymb]{revtex4-2}
\usepackage{graphicx}
\usepackage{dcolumn}
\usepackage{bm}
\usepackage{natbib}
\usepackage{color}
\usepackage{amsmath}
\usepackage[]{hyperref}
\hypersetup{
    colorlinks=true,
    allcolors = {blue},
    }
\usepackage[normalem]{ulem}
\newcommand{\remove}[1]{}

\newcommand{\add}[1]{\textcolor{black}{#1}}

\begin{document}

\title{Superconducting bistability in floating Al islands of hybrid Al/InAs nanowires}
\author{E.V.~Shpagina}\affiliation{Osipyan Institute of Solid State Physics, Russian Academy of
Sciences, 142432 Chernogolovka, Russian Federation}
\affiliation{National Research University Higher School of Economics, 20 Myasnitskaya Street, 101000 Moscow, Russian Federation}
\author{E.S.~Tikhonov}\affiliation{Osipyan Institute of Solid State Physics, Russian Academy of
Sciences, 142432 Chernogolovka, Russian Federation}
\affiliation{\add{Condensed-Matter Physics Laboratory, HSE University, Moscow 101000, Russia}}
\author{D.~Ruhstorfer}
\affiliation{Walter Schottky Institut, Physik Department, and Center for Nanotechnology and Nanomaterials, Technische Universit\"{a}t M\"{u}nchen, Am Coulombwall 4, Garching 85748, Germany}
\author{G.~Koblm\"{u}ller}
\affiliation{Walter Schottky Institut, Physik Department, and Center for Nanotechnology and Nanomaterials, Technische Universit\"{a}t M\"{u}nchen, Am Coulombwall 4, Garching 85748, Germany}
\author{V.S.~Khrapai}
\affiliation{Osipyan Institute of Solid State Physics, Russian Academy of
Sciences, 142432 Chernogolovka, Russian Federation}
\affiliation{National Research University Higher School of Economics, 20 Myasnitskaya Street, 101000 Moscow, Russian Federation}
\email{dick@issp.ac.ru}

\begin{abstract} 
We investigate a non-equilibrium aspect of the current-driven superconducting-normal phase transition in floating Al islands of epitaxial full-shell Al/InAs nanowires. Within a transition region discontinuous voltage jumps and hysteretic behaviour of the $I$-$V$ characteristics are observed, associated with the destruction and recovery of the superconducting order parameter in the island. The strength of the two features varies strongly in different devices in a mutually correlated way and can be suppressed by a small magnetic field. Numerical calculation explains this behaviour in terms of a tiny non-equilibrium correction to the electronic energy distribution at low energies. The experiment demonstrates a critical failure of a two-temperature non-equilibrium  model of the superconductor-normal transition in floating islands of hybrid nanowire devices. 
\end{abstract}

\maketitle

The search for Majorana zero modes in semiconducting nanowires (NWs) proximitized by a superconductor~\cite{lutchyn2010,oreg2010} brought a new generation of epitaxial hybrid devices~\citep{krogstrup2015,khan2020a,perla2021a}. \remove{As a side research direction, devices }\add{Beyond the Majorana research~\cite{SanHose2023,Paya2024a,Paya2024b,Deng2025}, devices }with the semiconducting core fully covered by the superconducting shell turned attractive for the studies of \remove{mesoscopic }flux-tunable superconductivity~\cite{vaitiekenas2020b} and its non-equilibrium aspects~\citep{ibabe2023b}. In the latter case, an interplay of quasiparticle injection, trapping and relaxation was found responsible for the suppression of superconductivity by transport current ($I$) in floating islands of Al/InAs NWs~\citep{shpagina2024,ibabe2024}.   

Non-equilibrium superconductivity is an interaction phenomenon, which manifests in the inter-dependence of the order parameter $\Delta$ and electronic energy distribution $f(E)$ (EED) in the parent superconductor~\citep{kopnin2001,keizer2006,snyman2009}. In the case of homogeneous energy-mode (heat-mode) excitation~\citep{tinkham2004}, the BCS (Bardeen-Cooper-Schrieffer) self-consistency equation can be expressed as:
\begin{equation}
\Delta = \frac{1}{\ln\left(2E_\mathrm{D}/\Delta_0\right)}\int_0^{E_\mathrm{D}} \mathrm{Im}(\sin\theta)\left[1-2f(E)\right]dE ,
    \label{BCS}
    \end{equation}    
where $E_\mathrm{D}$ is the Debye energy, $\Delta_0$ is the zero-temperature ($T$=0) order parameter and $\theta(E)$ is the energy-dependent complex-valued pairing angle in the Uzadel theory~\citep{anthore2003}. In a voltage ($V$) biased normal-superconductor-normal device and without relaxation, the non-equilibrium EED is double-step shaped and the superconductivity is suppressed for $|eV|\gtrsim\Delta_0$~\citep{keizer2006,snyman2009}. Moreover, the dependence $\Delta(V)$ predicted by the Eq.~(\ref{BCS}) in this case is bistable~\citep{keizer2006,snyman2009,moor2009,bobkova2013,kawamura2024,kawamura2025,bubis2021a}, which results in a hysteretic behavior of the $I$-$V$ characteristics in experiments~\citep{vercruyssen2012,shpagina2024}.

Transport across the floating superconducting islands in NW devices, that is the islands not connected to a quasiparticle reservoir, is a lively research topic~\citep{fu2010,vekris2022,wang2022b,liu2023,valentini2025}. Under the conditions of strong non-equilibrium this system represents a special case of mesoscopic superconductor. The finite $\Delta$ in the island survives up to much higher voltages $|eV|\gg \Delta_0$ thanks to a strong  relaxation of trapped quasiparticles, as observed recently in Refs.~\citep{shpagina2024,ibabe2024}. In these works, the suppression of the superconductivity is largely consistent with the assumption of the electronic temperature $T_\mathrm{e}$ elevated at increasing current, so that the order parameter follows the usual equilibrium dependence $\Delta(T_\mathrm{e})$. However, the observations of discontinuous voltage jumps, weak hysteresis and gate-voltage dependence of the critical Joule power~\citep{shpagina2024} indicate that the full local equilibrium is not reached.

In this work, we investigate deviations from the locally equilibrium EED in Al islands of epitaxial full-shell Al/InAs NWs. We observe that the key non-equilibrium features -- the discontinuous voltage jumps and the  $I$-$V$ hysteresis -- can be explained by a tiny admixture of non-thermalized low energy quasiparticles. Consistent with the experiment, in this scenario the strength of the hysteresis and the magnitude of the voltage jump are mutually correlated and both vanish in a finite magnetic field ($B$), as a result of smearing of the divergence of $\theta(E)$ near the gap edge. Our results shed light on the non-equilibrium aspect of the superconducting-normal transition in floating islands of Al/InAs NWs.

Our devices are individual InAs NWs with in-situ deposited full-shell Al. The growth is performed using molecular beam epitaxy via a selective area growth approach on SiO$_2$/Si(111) substrates~\cite{ruhstorfer2021,delgiudice2020,rudolph2013}. Typical NW diameter is about 160\,nm and the Al shell thickness is about 40\,nm. The NWs are manually transferred onto pre-patterned Cr/Au pads, with subsequent partial wet etching of the shell to form 0.85 to 2.2\,$\mathrm{\mu m}$ long Al islands \add{freely suspended above the substrate}. The processing is analogous to our previous work~\citep{shpagina2024}, and some of the devices used (D2 and D3) are the same as there. Here we mainly concentrate on the two samples with large (D1) and small (D2) hysteresis. All measurements are performed in a $^3$He insert with the sample immersed in liquid. \add{In some cases a positive gate voltage is used to minimize possible charging effects.}
\begin{figure}[t]
\begin{center}
\vspace{0mm}
  \includegraphics[width=1\linewidth]{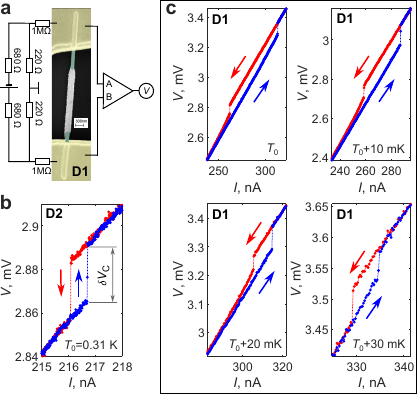}
\end{center}
  \caption{Non-equilibrium features on the $I-V$ curve. (a): false color scanning electron micrograph of the device D1 with a schematics of the symmetrized quasi-four-terminal current-biased measurement. (b),(c): $I-V$ curves in the device D2 with a relatively weak hysteresis (b) and in the device D1 with a relatively strong hysteresis (c). Shown are small current/voltage ranges next to the transition points for the two sweep directions, as indicated by blue and red colors and respective arrows. In panel (b) the magnitude of the voltage jump $\delta V_\mathrm{C}$ observed for the upward sweep in the device D2 is indicated. Sub-panels in (c) show the effect of small variations of the bath temperature on the hysteresis in the device D1. All data is taken in $B$=$0$. \add{For D1 back-gate and side-gate voltages are set to 15\,V, see SM~\citep{supplemental}. For D2 the back-gate voltage is 32\,V.} }
	\label{Fig1}
\end{figure}

Fig.~\ref{Fig1}a shows a false color scanning electron micrograph of the device D1 and the schematics of a quasi four-terminal symmetrized $I$-$V$ measurement. The central section of the NW is the Al island (grey), connected to the Au pads (yellow) via the  InAs core segments (green). Figs.~\ref{Fig1}b  and~\ref{Fig1}c show parts of the  $I$-$V$ curves ($I>$0, $B$=$0$) in the devices D2 and D1, respectively,  which detail the behavior near the superconducting-normal  transition in the Al island. At increasing $I$ (blue curves) the measured voltage exhibits a small upward jump marked with $\delta V_\mathrm{c}$ in Fig.~\ref{Fig1}b. On the backward sweep (red curves) a similar downward voltage jump occurs at a lower value of $I$, manifesting a hysteresis. At the lowest achievable bath temperature of $T_0\approx 0.31$\,K, the width of the hysteresis loop varies from $<1$\,nA upto $\sim50$\,nA in different devices. In devices with large hysteresis the behaviour is\remove{ slightly more complex} \add{richer}. We find that the width of the loop is very sensitive to a minor increase of the bath temperature.  \add{In addition, at the lowest $T_0$, the backward sweeps show a more complex two-stage return on the superconducting branch, note the difference between the red and blue lines below the downward voltage jump in} Fig.~\ref{Fig1}c. 
\add{Examples of the full $I$-$V$ curves and switching statistics are given in the Supplemental Materials (SM)~\citep{supplemental}.}
In the following we will investigate a mutual relation of $\delta V_\mathrm{c}$, the characteristic width of the hysteresis loop and the non-equilibrium EED in the Al island.
\begin{figure}[t!]
\begin{center}
\vspace{0mm}
  \includegraphics[width=1\linewidth]{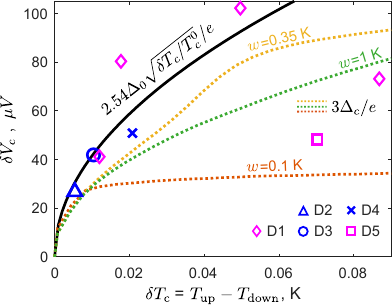}
\end{center}
  \caption{Correlation between the voltage jump and hysteresis. The magnitude of the voltage jump $\delta V_\mathrm{c}$ measured on the upward sweep (see Fig.~\ref{Fig1}b) is plotted as a function of the width $\delta T_\mathrm{c}$ of the hysteresis loop. $\delta T_\mathrm{c}$ is obtained from the difference of the electronic temperatures corresponding to the jumps on the upward and downward current sweeps, within the two-temperature model. The data from three devices with the weak hysteresis (D2-D4) and one device with the strong hysteresis (D1) are used, see legend. For the device D1 different symbols correspond to different bath temperatures from the sub-panels of Fig.~\ref{Fig1}c. Solid line is the theoretical estimate, dotted lines are the fits based on the model calculation of Fig.~\ref{Fig3}, see text. All experimental data is taken in $B$=$0$. \add{Gate-voltages for D1 and D2 are the same as in Fig.~\ref{Fig1}. For D5 the back-gate voltage is 32\,V. No gate voltage was used for D2 and D3.}} 
	\label{Fig2}
\end{figure}

 In spirit of the Joule spectroscopy approach~\citep{ibabe2023b}, we employ the two-temperature model and convert the position of the voltage jump into the electronic temperature of the Al island. We use a relation $P_\mathrm{J}/2=\mathcal{V}_\mathrm{Al}\Sigma_\mathrm{e-ph}\left(T_\mathrm{e}^5-T_\mathrm{0}^5\right)$, where $P_\mathrm{J}=IV$ is the total released Joule power, $\mathcal{V}_\mathrm{Al}$ is the island volume and $\Sigma_\mathrm{e-ph} = 4.8 \,\mathrm{nW}\mu\mathrm{m}^{-3}\mathrm{K}^{-5}$ is the electron-phonon cooling power~\citep{shpagina2024}. In this way the transition temperatures 
$T_\mathrm{up}$ and $T_\mathrm{down}$ for the two sweep directions are obtained and the width of the hysteresis loop is expressed in terms of their difference $\delta T_\mathrm{c}$= $T_\mathrm{up}-T_\mathrm{down}$. In Fig.~\ref{Fig2} we collect the data $\delta V_\mathrm{c}(\delta T_\mathrm{c})$ from different devices. The overall trend is the increase of $\delta V_\mathrm{c}$ accompanied by the increase of the hysteresis, both among different devices and for the device D1 measured at slightly different bath temperatures, see Fig.~\ref{Fig1}c. This behaviour can be discussed quantitatively, as explained below. 

Typical measured $\delta V_\mathrm{c}$ corresponds to a change of the device resistance by $\delta V_\mathrm{c}/I = 100-300\,\Omega$, which is much higher than the normal resistance of the Al ($\sim 1\,\Omega$) and much lower than the InAs NW resistance ($\sim 10\,\mathrm{k}\Omega$). We associate the voltage jump with the change of the resistances of the two InAs NW sections adjacent to the island, which occurs in response to the collapse of the superconductivity in the Al. The resistance between the diffusive metallic wire and the superconductor is diminished by the Andreev reflection process, which results in the excess current ($I_\mathrm{exc}$) as compared to the normal state~\citep{artemenko1979,bardas1997,brinkman2003}. In the relevant limit of $|eV|\gg k_\mathrm{B}T,\,\Delta$ the excess current is given by the expression~\citep{artemenko1979} $I_\mathrm{exc}R_\mathrm{N} \approx 0.73\Delta/\add{|e|}$, where  $R_\mathrm{N}$ is the wire's normal resistance. For our devices we may thus relate the observed voltage jump to the jump of the order parameter from a finite value $\Delta_\mathrm{c}$ to zero via $\delta V_\mathrm{c}\approx 1.46\Delta_\mathrm{c}/\add{|e|}$. Taking $\delta T_\mathrm{c}$ as a measure of the detuning from the superconducting critical temperature and using a standard BCS relation, we estimate $\Delta_\mathrm{c} \approx 1.74\Delta_0\sqrt{\delta T_\mathrm{c}/T^0_\mathrm{c}}$, where $T^0_\mathrm{c}\approx1.23$\,K in our devices~\citep{shpagina2024} and $\Delta_0\approx1.76 k_\mathrm{B}T^0_\mathrm{c}\approx 187\,\mu$eV is the BCS gap at $T$=0. The obtained dependence $\delta V_\mathrm{c}\left(\delta T_\mathrm{c}\right)$ is shown in Fig.~\ref{Fig2} by the solid line and captures well the observed experimental behaviour. This agreement ensures that similar physics takes place in all devices irrespective of the size of the hysteresis loop. 
\add{Note, however, that the above estimates are not self-consistent, since the very existence of the hysteresis and voltage jumps remains inexplicable within the equilibrium BCS theory. Below we present an effort of self-consistent non-equilibrium treatment.}
\begin{figure}[t!]
\begin{center}
\vspace{0mm}
  \includegraphics[width=1\linewidth]{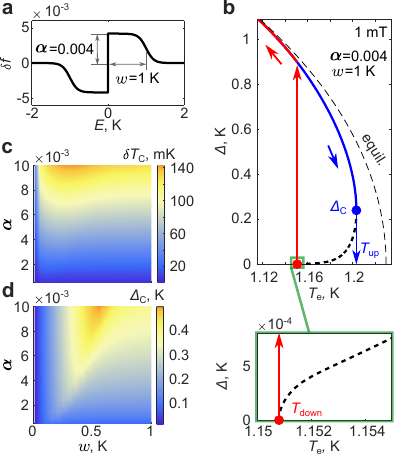}
\end{center}
  \caption{Model of the quasiparticle non-equilibrium. (a): non-equilibrium correction $\delta f(E)$ to the EED in the Al island, controlled with the two parameters $\alpha$ and $w$. (b): dependence of the order parameter on the electronic temperature, which controls the equilibrium part of the EED with the fixed $\delta f(E)$, see the indicated $\alpha$ and $w$. The thick solid lines show the stable branch of  $\Delta(T_\mathrm{e})$ for two sweep directions indicated with arrows. The discontinuous jump of the order parameter on the upward sweep is marked as $\Delta_\mathrm{c}$. The thick dashed line is the unstable branch of  $\Delta(T_\mathrm{e})$. The thin dashed line is the equilibrium dependence calculated with $\delta f(E)\equiv0$. (c): colorscale plot of the width of the hysteresis loop $\delta T_\mathrm{c}$ as a function of  $\alpha$ and $w$. (d): colorscale plot of the jump  $\Delta_\mathrm{c}$ of the order parameter at the high temperature end of the hysteresis loop as a function of  $\alpha$ and $w$. All data is calculated for $B=1\,$mT.  } 
	\label{Fig3}
\end{figure}

The bistability of the order parameter is a signature of quasiparticle non-equilibrium, which plays crucial role near the phase transition. We explore the role of non-equilibrium quasiparticles using the Eq.~(\ref{BCS}) with the EED  $f(E) = f_0(E,T_\mathrm{e})+\delta f(E)$, where the first term is the Fermi-Dirac distribution with the given electronic temperature and the second term is a small non-equilibrium correction. We guess $\delta f(E)$
in the form of step function with the shape controlled with two parameters: amplitude $\alpha$ and width $w$, see Fig.~\ref{Fig3}a and\remove{ Supplemental Materials (SM)} \add{SM}~\citep{supplemental} for the expression. This choice allows to capture the effect of non-thermalized quasiparticles in the Al island, both electrons ($E > 0$) and holes ($E < 0$). Such quasiparticles originate in the course of injection
of hot carriers with $|E|\sim |eV/2|$ from the normal leads and their relaxation in the island. Fig.~\ref{Fig3}b shows the calculated dependence $\Delta (T_\mathrm{e})$ for the case of $\alpha$=0.004 and $w$=1\,K (thick lines). Here we set a small  magnetic field of $B=1\,$mT to improve the convergence of the numerical integration. The effect of the $B$-field is to introduce finite depairing and  remove a singularity in the dependence $\theta(E)$ in the Eq.~(\ref{BCS}). Sufficiently below $T_\mathrm{c}$ the order parameter coincides with the equilibrium dependence (thin dashed line), but deviates downward from it at increasing $T_\mathrm{e}$. In the range  $T_\mathrm{down}$ $<T_\mathrm{e}$ $<T_\mathrm{up}$ two solutions for $\Delta$ coexist, which we interpret  in terms of a bistability of the order parameter~\citep{keizer2006,snyman2009,shpagina2024}. The upper stable branch (blue line) corresponds to the decrease of $\Delta$ at increasing temperature and explains a discontinuous jump of the order parameter from $\Delta$=$\Delta_\mathrm{c}$ to zero when  superconductivity collapses at $T_\mathrm{e}$=$T_\mathrm{up}$. The lower branch (thick dashed line) is unstable, so that at decreasing $T_\mathrm{e}$ the superconductivity is restored only at $T_\mathrm{e}$=$T_\mathrm{down}$ with another jump of the order parameter (red line). The dependencies of $\delta T_\mathrm{c}$ and $\Delta_\mathrm{c}$ on $\alpha$ and $w$ are shown on color scale plots of Figs.~\ref{Fig3}c and ~\ref{Fig3}d. We observe that $\delta T_\mathrm{c}$ primarily depends on the parameter $\alpha$ (Figs.~\ref{Fig3}c), whereas $\Delta_\mathrm{c}$ is sensitive to both $\alpha$ and $w$ (Figs.~\ref{Fig3}d). Using fixed $w$ and varying $\alpha$ we calculate the dependence $\Delta_\mathrm{c}(\delta T_\mathrm{c})$ and fit the data of Fig.~\ref{Fig2}, see the dotted lines obtained for\remove{ the two} \add{several} indicated values of $w$. Here, reasonable fits are obtained for $\delta V_\mathrm{c} = 3\Delta_\mathrm{c}/\add{|e|}$ meaning that the model captures the experiment but underestimates the magnitude of the voltage jump. In addition, the downward voltage jumps observed on the $I$-$V$ curves of Fig.~\ref{Fig1} are comparable to the upward jumps, whereas the model predicts them to be considerably stronger (see the upper panel of Fig.~\ref{Fig3}b). \add{This discrepancy comes from the fact that our model predicts the order parameter in the island to be fully restored at $T_\mathrm{e}$=$T_\mathrm{down}$, whereas the behaviour observed on the backward sweeps in Fig.~\ref{Fig1}c suggests a two-stage process with an intermediate superconducting phase realized just below the downward voltage jump. Possible candidates for such a phase are a bimodal superconducting state observed earlier in metallic devices~\citep{vercruyssen2012} or a non-equilibrium Fulde-Ferrell-Larkin-Ovchinnikov type state~\cite{moor2009,bobkova2013,kawamura2024,kawamura2025}.} \remove{This discrepancy may reflect a more complex behaviour at the transition from the normal to the superconducting state, as observed earlier in metallic devices~\citep{vercruyssen2012}.}

\begin{figure}[t!]
\begin{center}
\vspace{0mm}
  \includegraphics[width=1\linewidth]{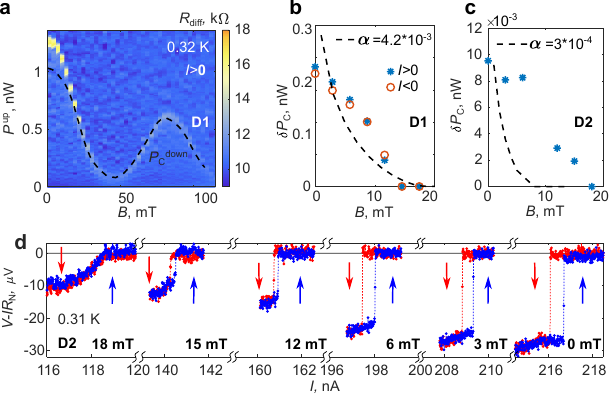}
\end{center}
  \caption{Impact of the $B$-field. (a): color scale plot of the numerically calculated differential resistance of the device D1 as a function of $P_\mathrm{J}$ and $B$  (upward sweep). Dashed line indicates the position of the $R_\mathrm{diff}$ maximum extracted from similar data for the downward sweep. (b): the width of the hysteresis loop in units of the Joule power for the device D1 (symbols) and the model fit with the indicated $\alpha$ and $w=1\,$K (dashed line). (c): the same for the device D2. (d): magnified view of the transition region on the $I$-$V$ curve for a set of $B$-fields in device D2.  A linear background corresponding to the slope in the normal state at high currents is subtracted from the data. Two sweep directions are indicated with separate colors and arrows. \add{Gate voltages are the same as in Fig.~\ref{Fig1}.}} 
	\label{Fig4}
\end{figure}
As a final step, we investigate the role of the $B$-field on the non-equilibrium phase transition. Fig.~\ref{Fig4}a shows a color-scale plot of the differential resistance ($R_\mathrm{diff}$) of the device D1 obtained from the numerical derivative of the $I$-$V$ curves.  $R_\mathrm{diff}$ is plotted as a function of $P_\mathrm{J}$ and the $B$-field applied along the NW. This data is obtained for the upward current sweeps (direction from $I$=$0$). The voltage jumps translate into $R_\mathrm{diff}$ peaks and form a yellow line, whose position oscillates with $B$ owing to the Little-Parks (LP) effect. For comparison, a position of a similar feature on the downward sweeps (direction to $I$=$0$) is shown by the dashed line. Fig.~\ref{Fig4}a demonstrates
that the $R_\mathrm{diff}$ peak becomes fainter at increasing the $B$-field because of the smearing of the voltage jumps~\citep{shpagina2024} 
and the hysteresis observed in $B$=0 disappears already
in $B\approx15\,$mT. The same qualitative  behaviour is found in the device D2, as shown in Fig.~\ref{Fig4}d. Here we plot the $I$-$V$ curves with the subtracted background for a set of $B$-fields within the first LP lobe. The hysteresis shrinks and the voltage jumps gradually diminish before both disappear in $B=18$\,mT. The dependence of the hysteresis, now expressed in terms of the Joule power $\delta P_\mathrm{c}$, on the $B$-field is plotted in Figs.~\ref{Fig4}b and ~\ref{Fig4}c for the two devices (symbols). The $B$-field suppresses the hysteresis in a similar fashion, in spite of the very different magnitude of the effect in the devices D1 and D2. Our model captures this behaviour qualitatively, see the dashed lines, predicting even a somewhat stronger $B$-field effect. Similar to the effect of life-time broadening~\citep{snyman2009}, the impact of the $B$-field comes from the depairing factor $\Gamma\propto B^2$, which removes the singularity in $\mathrm{Im}(\sin\theta)$ in the Eq.~(\ref{BCS}). This effect is strongest in the range of $|E|\leq 1\,$K, see the SM~\citep{supplemental} for the details, supporting our conjecture that the observed behaviour originates from the low energy non-equilibrium quasiparticles.

The above analysis strongly suggests that the current driven superconducting-normal phase transition in floating islands of Al/InAs NWs is a non-equilibrium effect resulting from a tiny deviation of the quasiparticle EED from the local equilibrium. The deviation is on the order of $\delta f/f_0\sim 10^{-2}$ in devices with strong hysteresis (like D1) and $\delta f/f_0\sim 10^{-3}$ in devices with weak hysteresis (like D2). The smallness of $\delta f$ explains why the two-temperature model was found largely consistent with the data in the previous  experiments~\citep{shpagina2024,ibabe2024}. The key argument of the two-temperature model is large quasiparticle dwell time $\tau_\mathrm{dwell}$ as compared to the electron-phonon relaxation time $\tau_\mathrm{e-ph}$. Even $\tau_\mathrm{dwell}\gg\tau_\mathrm{e-ph}$, however, cannot guarantee that $\delta f (E)\equiv0$ in the steady state. The fact that the ratio  $\delta f/f_0$ strongly fluctuates among the different samples may not seem very surprising for such tiny $\delta f$, although we could not identify a systematic dependence of the width of the hysteresis loop on the specific sample parameter. A hint that the phonon degree of freedom can play a non-trivial role comes from the fact that $k_\mathrm{B}T_0\sim\Delta_\mathrm{c}$, i.e. the typical phonon energy, given by the bath temperature, is comparable to the energy gap. Hence, for the quasiparticles near the gap edge the relaxation slows down as compared to the equilibrium case near the $T_\mathrm{c}$ and the two-temperature model is more difficult to justify.      

In summary, we investigated the non-equilibrium aspect of the current driven superconducting-normal phase transition in floating Al islands of the epitaxial Al/InAs NWs. The transition is accompanied by voltage jumps and hysteresis associated
with the destruction/recovery of the superconducting proximity effect in the InAs core. A small magnetic field suppresses these features and makes the transition continuous. The model based on the assumption of small non-equilibrium correction to the low-energy quasiparticle population captures the experimental observations correctly. Our results demonstrate the critical importance of the self-consistent treatment of the order parameter, quasiparticle population and relaxation in floating superconducting islands of hybrid NW devices out of equilibrium. 
\remove{The present experiment may also be useful as a step towards the realization of Fulde-Ferrell-Larkin-Ovchinnikov type non-equilibrium superconductivity in hybrid devices [19-22].}

We acknowledge valuable discussions with A.M. Bobkov and A.A. Golubov. \add{The work of E.S.T. was partially supported by the Basic Research Program of HSE.}


%

\end{document}


\title{Superconducting bistability in floating Al islands of hybrid Al/InAs nanowires. Supplemental Materials.}
\author{E.V.~Shpagina}\affiliation{Osipyan Institute of Solid State Physics, Russian Academy of
Sciences, 142432 Chernogolovka, Russian Federation}
\affiliation{National Research University Higher School of Economics, 20 Myasnitskaya Street, 101000 Moscow, Russian Federation}
\author{E.S.~Tikhonov}\affiliation{Osipyan Institute of Solid State Physics, Russian Academy of
Sciences, 142432 Chernogolovka, Russian Federation}
\affiliation{\add{Condensed-Matter Physics Laboratory, HSE University, Moscow 101000, Russia}}
\author{D.~Ruhstorfer}
\affiliation{Walter Schottky Institut, Physik Department, and Center for Nanotechnology and Nanomaterials, Technische Universit\"{a}t M\"{u}nchen, Am Coulombwall 4, Garching 85748, Germany}
\author{G.~Koblm\"{u}ller}
\affiliation{Walter Schottky Institut, Physik Department, and Center for Nanotechnology and Nanomaterials, Technische Universit\"{a}t M\"{u}nchen, Am Coulombwall 4, Garching 85748, Germany}
\author{V.S.~Khrapai}
\affiliation{Osipyan Institute of Solid State Physics, Russian Academy of
Sciences, 142432 Chernogolovka, Russian Federation}
\affiliation{National Research University Higher School of Economics, 20 Myasnitskaya Street, 101000 Moscow, Russian Federation}
\email{dick@issp.ac.ru}

\maketitle
\section{Devices}
\begin{figure}
	\begin{center}
		\vspace{0mm}
		\includegraphics[width=1\linewidth]{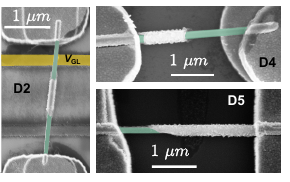}
	\end{center}
	\caption{ Scanning electron micrographs of the devices \add{D2,} D4 and D5. \add{The green color shows areas with bare InAs. The $V_{GL}$ gate is shown in yellow. }  } 
	\label{Fig.S1}
\end{figure}

The series of devices discussed here includes devices D2 and D3, which were previously discussed in~\citep{shpagina2024} and were referred to as IIE and IIA, respectively. Scanning electron micrographs of the devices D4 and D5 are shown in Fig.~\ref{Fig.S1}. \add{This figure also includes a micrograph of device D2, equipped with the metallic gate to which the voltage $V_{GL}$ was applied in the measurements described in the main text (false color). The gate is  formed by a 2/25 nm Cr/Au bilayer deposited on the substrate and spatially separated from the nanowire. All nanowires are placed on pre-fabricated Cr/Au pads with thicknesses of 2/150 nm, respectively, which suspend the nanowire along with the superconducting island above the substrate.}

\section{Heat balance  equation}
Following the Joule spectroscopy approach~\citep{shpagina2024,ibabe2023b}, we utilize a two-temperature model to convert the voltage jump position into the electronic temperature of the Al island ($T_\mathrm{e}$). The heat balance equation is described by 

\begin{align}
	P_\mathrm{J}/2 = \mathcal{V}_\mathrm{Al} \Sigma_\mathrm{e-ph}(T_\mathrm{e}^5 - T_0^5), \label{Joule}
\end{align}

where $P_\mathrm{J} = IV$ represents the total Joule power, half of which is dissipated in the S-island. $\mathcal{V}_\mathrm{Al}$ is the island volume, and $\Sigma_\mathrm{e-ph}$ is the electron-phonon cooling power. $T_0$ is the  bath temperature. The dependence of superconducting critical temperature $T_\mathrm{c}$ on the magnetic field $B$ is found from the solution to the well-known Abrikosov-Gorkov equation:
\begin{align}
	\ln \left( \frac{T_\mathrm{c}}{T_\mathrm{c}^0} \right) = \Psi \left(\frac12 \right) - \Psi \left(\frac12 + \frac{\Gamma(B,n) }{2\pi T_\mathrm{c}} \right) \label{AG_eq}
\end{align}
where $\Psi$ is the digamma function, $T_\mathrm{c}^0\approx 1.23$\,K is the critical temperature in zero $B-$field.

The depairing factor $\Gamma$ is determined as a function of the winding number ($n$) and magnetic field, decomposed into parallel and perpendicular components $\Gamma(B,n) = \Gamma_{\perp} + \Gamma_{\parallel}$. Following established models for full-shell nanowires ~\citep{vaitiekenas2020b,ibabe2023b}, we describe the epitaxial Al shell using a hollow cylinder approximation under arbitrary magnetic fields, incorporating finite thickness ($t$). Assuming the superconducting film thickness remains below the coherence length ($t<\xi_0$), the depairing factor reduces to the radius-averaged square of the superconducting velocity~\citep{anthore2003}. As explained in~\citep{shpagina2024}, we then derive the expression for components of depairing factor 
\begin{align*}
	\Gamma_{\parallel} &= \frac{\hbar D}{2 \rho_+^2}
	\left[
	\left(n-\frac{\Phi}{\Phi_0}\right)^2
	+ n^2 
	\left(
	\frac{\rho_+^2}{\rho_-^2}
	\ln \frac{\rho_\mathrm{i}+t}{\rho_\mathrm{i}} - 1
	\right)
	\right], \\
	\Gamma_{\perp} &= \frac{\hbar D}{\Phi_0^2} (\pi B_\perp \rho_+)^2
\end{align*}
where $D=\xi_0^2 \Delta_0/\hbar$ is the diffusion coefficient, expressed in terms of the order parameter at a zero electron temperature  ($\Delta_0=2.1$ K for aluminum), $\rho_\pm^2 = ((\rho_\mathrm{i}+t)^2 \pm \rho_\mathrm{i}^2)/2 $, $\Phi=\pi \rho_+^2 B_\parallel$ and $n=0,1,2$ is the number of the Little-Parks lobe, $\Phi_0 = h/2e$. $\rho_i$ is the radius of the InAs core. In the $n=0$ loop the total depairing factor depends on the $B$-field as $\Gamma \propto B^2$.

\begin{figure}
	\begin{center}
		\vspace{0mm}
		\includegraphics[width=0.9\linewidth]{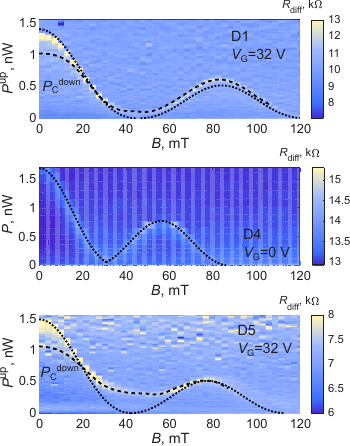}
	\end{center}
	\caption{ Color-scale plots of $R_\mathrm{diff} (B, V )$ for the devices not presented in Ref.~\citep{shpagina2024} (upward sweeps, current from zero). The gate voltage ($V_\mathrm{G}$) is indicated in the legend. Dashed lines indicate the critical power when the current sweep is directed to zero (downward sweep, $P_C^{down}$). Dotted lines represent the fit of the critical power calculated from the heat balance  equation~(\ref{Joule}) using parameters listed in the TABLE~\ref{table}.} 
	\label{Fig.S2}
\end{figure}

The fitting procedure involves determining $\mathcal{V}_\mathrm{Al}$ and the misalignment angle $\alpha$  (between the $B$-field and the NW axis) from the color-plot $R_\mathrm{diff}(B, I)$ and extracting $\Sigma_\mathrm{e-ph}$ from the critical Joule power in $B=0$. The results of the fitting and the corresponding parameters extracted for thes previously unpublished \add{devices} are shown in Fig.~\ref{Fig.S2} and listed in TABLE~\ref{table}. In all devices the shell thickness is $t=42$~nm and the superconducting coherence length is $\xi_0=156$~nm. The value of $\Sigma_\mathrm{e-ph}$ obtained for the device D4 is about a factor of 2 higher than in all other \add{devices}. This discrepancy is most likely  a mistake, although we could not identify its origin.

\begin{table}[t!]
	\centering 
	\caption{ Parameters of the devices. }          
	\begin{tabular}{|c|c|c|c|c|c|c|}\hline Device &  $\rho_\mathrm{i}$, nm & $L$, $\mu$m & $\mathcal{V}_\mathrm{Al}$, $\mu$m$^3$ & $\alpha$,$^o$ & $\Sigma_\mathrm{e-ph}$, $\frac{nW}{ \mu m^{3} K^{5} }$ \\\hline
		D1 & 65 & 2.2	& 0.049 & 0  & 5 \\
		D4 & 83 & 1.1	& 0.030 & 8 & 10 \\
		D5 & 68 & 2.2   & 0.052 & 5 & 5 \\\hline
	\end{tabular} \label{table}
\end{table}

\section{ Features on the $I$-$V$ curve }
\begin{figure}[h!]
	\begin{center}
		\vspace{0mm}
		\includegraphics[width=1\linewidth]{ 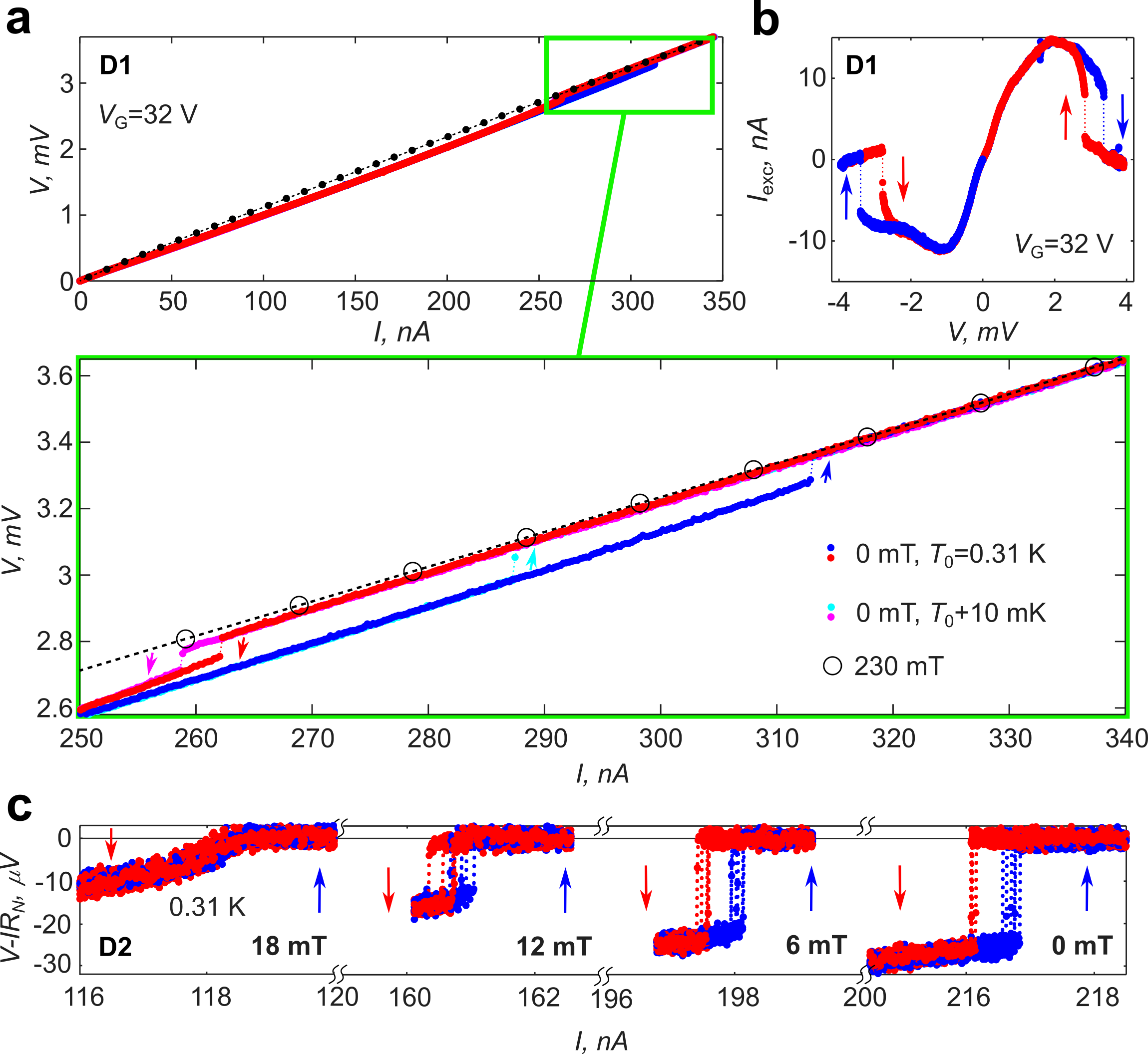}
	\end{center}
	\caption{ 
	\add{
	Non-equilibrium features in the $I–V$ characteristics.  
	(a): $I–V$ curves for D1, measured in the following order. Upward (blue) and downward (red) sweeps at temperature $T_0$, black curve measured in $B=$230\,mT, and upward/downward sweeps (light blue/magenta) at $T_0 + 10$ mK. Zero-field data are the same as in Fig.~1c of the main text. A magnified view of the hysteresis is shown nearby. 
	(b): Excess current as a function of bias voltage, extracted from panel (a), calculated as the difference between currents with and without the applied magnetic field at temperature $T_0$.  
	(c): Statistics of voltage jumps  measured in D2 in several $B$-field values supplementing the data of Fig.~4d of the main text. }	  } 
	\label{Fig.S3}
\end{figure}
\begin{figure*}
	\begin{center}
		\vspace{0mm}
		\includegraphics[width=1\linewidth]{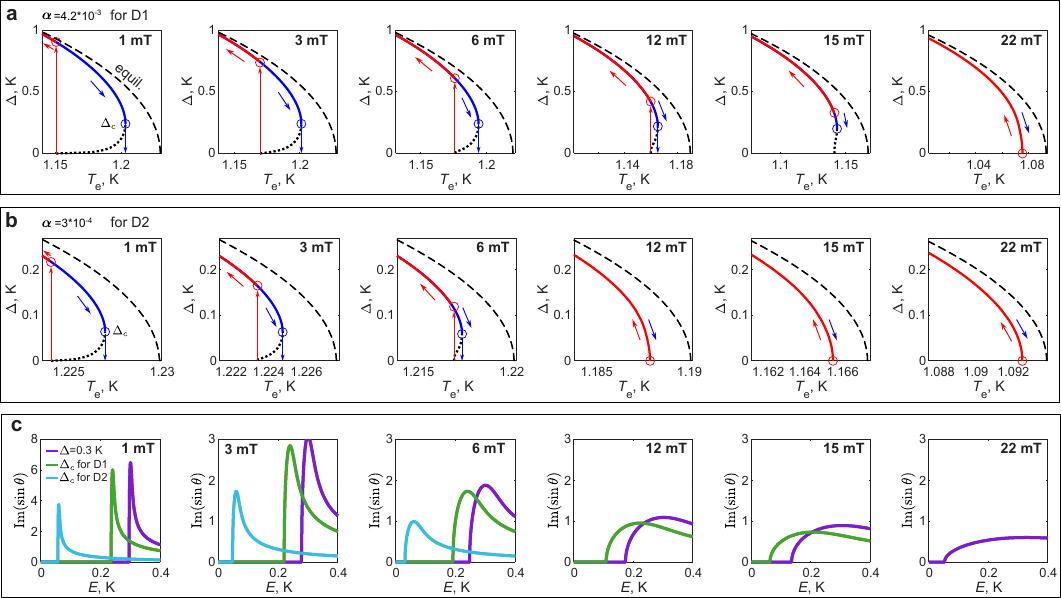}
	\end{center}
	\caption{ (a), (b): Calculated dependencies $\Delta(T_\mathrm{e})$ in equilibrium (dashed lines) and with a non-equilibrium correction (solid lines) for the devices D1 and D2. Dotted lines indicate unstable branches of the solution. The values of $\alpha$ used are given in the legends, $w=1\,$K. (c):  $\mathrm{Im}(\sin\theta)$ as a function of energy, calculated for the three values of $\Delta$ indicated in the legend. Each sub-panel is calculated in a given $B$-field indicated in the legend.} 
	\label{Fig.S4}
\end{figure*}

\add{This section supplements the $I$-$V$ characteristics presented in the main text with an extended dataset. The data from Figure~1c of the main text were acquired over a broader range of bias currents, including both negative and positive values. Between the $I$-$V$ measurements at $T_0$ and ($T_0+10$ mK), on the same day we took a measurement of the $I$-$V$ curve in a 230~mT magnetic field, which is high enough to completely suppress the superconductivity. Figure~\ref{Fig.S3}a displays the $I$-$V$ characteristics for $I>0$, comparing data measured with and without the magnetic field at the two base temperatures $T_0$ and $(T_0 + 10)$ mK. The region of hysteresis is detailed in a separate panel below. Figure \ref{Fig.S3}b shows the excess current as a function of bias voltage found from $I_{exc}(V) = I_N(V) - I(V)$. Here $I_N(V)$ is the 230\,mT trace and $I(V)$ is the $B$=0 trace at $T_0$. It is seen that $I_{exc}$ is a non-monotonic function of $V$. Initial growth of the $|I_{exc}|$ at increasing $|V|$ is followed by a decrease above $|V|\gtrsim 2$\,mV. This decrease is a result of diminishing order parameter $\Delta$ in the island caused by the increase of the electronic temperature $T_e$, see also below. Note that in the normal state at high currents the $B$=0 and $B$=230~mT $I$-$V$ curves do not coincide exactly, which may be caused by a tiny orbital effect of the $B$-field in InAs. This discrepancy gives rise in artificial $I_{exc}\neq0$ when the superconductivity is suppresses in $B=0$.}

\add{The position of the voltage jumps exhibit slight random fluctuations across consecutive $I$–$V$ sweeps. This effect is most pronounced in devices with small hysteresis, as shown in Fig.~\ref{Fig.S3}c. This data extends Fig.~3c from the main text, where only one pair of consecutive sweep-up and sweep-down curves for each $B$-field were shown. Fig.~\ref{Fig.S3}c shows five consecutive pairs of sweep-up/sweep-down curves measured for each $B$-field value. While each up–down pair shows clear hysteresis, its position and width fluctuate. Such fluctuations can be comparable to the typical width of the hysteresis, see the  $B$=12\,mT trace.
}

\section{Order parameter}
\begin{figure}
	\begin{center}
		\vspace{0mm}
		\includegraphics[width=1\linewidth]{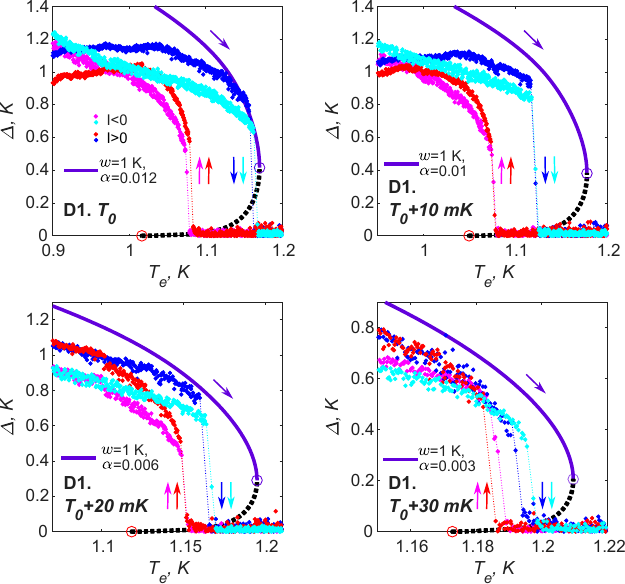}
	\end{center}
	\caption{  
		\add{Order parameter in the hysteresis region. A comparison of theory and experiment. Experimental data for the device D1 (from Fig. 1c in the main text) are shown for the two polarities of the bias and for the two sweep directions, see legend. The procedure used to extract $\Delta(T_e)$ from the $I$-$V$ curves is described in the text.  The theoretical up branch (purple) is calculated for $B=$1\,mT  using $w=1$\,K and varying $alpha$, as indicated in the legends. 
	}} 
	\label{Fig.S5}
\end{figure}

In our theoretical analysis, we employ the Usadel formalism, as described in Ref. ~\citep{anthore2003}, and incorporate the non-equilibrium electronic energy distribution (EED). We consider the non-equilibrium EED $f(E)$ in the island as a Fermi-Dirac distribution $f_0(E)$ with a small non-equilibrium correction $\delta f(E)$, which generates a small number of extra low-energy quasiparticles at low energies (typically $|E| \leq 1$ K):
\begin{align}
	f(E) =  f_0(E) + \delta f(E)
\end{align}

We take the correction in the form:
\begin{align}
	\delta f(E>0) =  \frac{\alpha}{ \exp \frac{(E-w)}{w/10} + 1 };\,\, \delta f(-E) = -\delta f(E)
\end{align}
where $\alpha$ represents the fraction of non-equilibrium quasiparticles and $w$ defines their energy range. Here $E>0$ corresponds to electrons and $E<0$ to holes. The parameter $\alpha$ is on the order of $10^{-4}-10^{-3}$ and $w\leq 1\,$K.

The order parameter $\Delta$ is related to the EED via the Bardeen-Cooper-Schrieffer (BCS) self-consistency equation:
\begin{equation}
	\Delta = \frac{1}{\ln\left(2E_\mathrm{D}/\Delta_0\right)}\int_0^{E_\mathrm{D}} \mathrm{Im}(\sin\theta)\left[1-2f(E)\right]dE ,
	\label{BCS}
\end{equation}    
where $E_\mathrm{D}$ is the Debye energy, $\Delta_0$ is the zero-temperature ($T$=0) order parameter and $\theta(E)$ is the energy-dependent complex-valued pairing angle in the Uzadel theory. For the sake of faster convergence we used $E_\mathrm{D}=20\,$K in the non-equilibrium calculation, which is enough for the purpose of this study. 

The pairing angle is determined from  the Usadel equation~\citep{anthore2003}:
 \begin{align}
	E + i \Gamma \cos \theta = i \Delta \frac{\cos \theta}{\sin \theta} \label{uzadel}
\end{align}
The calculation is performed as follows. First, for a given $B$-field we find $\Gamma$. Next, for a given $\Delta$ we find the spectrum of the superconductor from a solution of the Eq.~(\ref{uzadel}). Finally, for each $\alpha$ and $w$ we solve the Eq.~(\ref{BCS}) to find the dependence $\Delta(T_\mathrm{e})$.

The results of the calculation are shown in Fig.~\ref{Fig.S4} for a set of the $B$-field values. Panels (a) and (b) show the dependencies $\Delta(T_\mathrm{e})$ obtained for the devices D1 and D2, respectively. The used values of $\alpha$ for each device are displayed in the legend and $w=1\,$K is the same. Panel (c) demonstrate the energy dependence of the quantity $\mathrm{Im}(\sin\theta)$ for a fixed $\Delta = 0.3\,$K and for a critical value of the order parameter $\Delta=\Delta_\mathrm{c}$ in both devices. In both devices we find a bistable behaviour near the superconducting-normal non-equilibrium phase transition. The bistability is suppressed in moderate $B$-fields as a result of smearing of the singularity in the dependence $\theta(E)$ near the gap edge. The data of Fig.~\ref{Fig.S4} were used in  Fig.~4 of the main text.

\add{To enable a direct comparison between the theoretical dependence $\Delta(T_e)$ and the experimental data, we use the theory of excess current developed by Artemenko et al. in Ref.~\cite{Artemenko1979}. This work provides the relation between the $\Delta$ and $I_{exc}$ in an NS junction. We adapt their results in the limit $|eV|\gg \Delta, k_B T$ to our NSN devices, which yields the expression:
	\begin{align}
		\Delta \approx \left|eI_{exc}\right|R_N \left( \frac{\pi^2}{4}-1 \right)^{-1}, \label{Delta_eq}
	\end{align}
	where $R_N$ is the normal-state resistance of the device (extracted from the Fig.~\ref{Fig.S3}a). Applying this analysis to the data for D1 from Fig.~1c of the main text and using the Eq.~(\ref{Joule}) to convert the voltage bias in $T_e$ gives Fig.~\ref{Fig.S5}. The experimental dependencies $\Delta(T_e)$ are shown with symbols for the two bias polarities and two current sweep directions (see legend). In order to ensure  $I_{exc}=0$ (and thus $\Delta =0$) in the normal state additional linear trend was subtracted from the dependencies $I_{exc}(V)$. $\Delta(T_e)$ obtained in this way is almost independent of the bias polarity in the vicinity of $T_c$. As expected, the hysteresis and discontinuous jumps of the order parameter are observed. Purple solid lines and black dashed lines demonstrate the results of numerical calculation of $\Delta(T_e)$ with fixed $w=1$\,K and varying $\alpha$, respectively, for the stable and unstable branches. As already mentioned in the main text, the calculation is capable to adequately describe the width of the hysteresis loop, but predicts substantially lower values of the jump $\Delta_c$ of the order parameter (purple symbols). }



%